\documentstyle[multicol,prl,aps,epsf]{revtex}

\begin{document}

\title{Suppression of Giant Magnetoresistance by
a superconducting contact}
\author{F. Taddei\thanks{e-mail: f.taddei@lancaster.ac.uk},}
\address{School of Physics and Chemistry, Lancaster University,
Lancaster, LA1 4YB, UK}
\author{S. Sanvito\thanks{e-mail: sanvito@dera.gov.uk},}
\address{School of Physics and Chemistry, Lancaster University,
Lancaster, LA1 4YB, UK and \\
DERA, Electronic and Optical Materials Centre,
Malvern, Worcs. WR14 3PS UK}
\author{J.H.Jefferson,}
\address{DERA, Electronic and Optical Materials Centre,
Malvern, Worcs. WR14 3PS UK}
\author{C.J. Lambert\thanks{e-mail:c.lambert@lancaster.ac.uk}}
\address{School of Physics and Chemistry, Lancaster University,
Lancaster, LA1 4YB, UK}
\date{\today}
\maketitle

\begin{abstract}
We predict that current perpendicular to the plane (CPP) giant
magnetoresistance (GMR) in a phase-coherent magnetic multilayer
is suppressed when one of the contacts is superconducting.
This is a consequence of a superconductivity-induced magneto-resistive
(SMR) effect, whereby the conductance of the ferromagnetically aligned state is 
drastically reduced by superconductivity.
To demonstrate this effect, we compute the GMR ratio of clean (Cu/Co)$_n$Cu
and (Cu/Co)$_n$Pb multilayers, described by an {\it ab-initio} spd
tight binding Hamiltonian. By analyzing a simpler model with
two orbitals per site, we also show that the suppression survives in the
presence of elastic scattering by impurities. 
\end{abstract}

\begin{multicols}{2}



During the past decade, electronic properties of hybrid nanostructures have 
been widely studied, both from a fundamental point of view and for their 
potential applications. At a fundamental level, new physics associated with
such structures arises from the proximity of two electronic ground states with 
different correlations, which can reveal novel scattering processes not 
apparent in the separate materials. For example normal-superconducting hybrids,
formed when a normal (N) metal or semiconductor is placed in contact with
a superconductor (S), have been shown to exhibit a range of unique 
transport phenomena \cite{colin,karl} associated with Andreev scattering
at the N-S boundary.
Similarly ferromagnetic (F)-normal multilayers and spin-valves exhibit 
magnetoresistance properties \cite{gijs,prinz} associated with
spin-filtering by the F-layers. Perhaps the most spectacular of these is 
giant magnetoresistance (GMR) in magnetic multilayers, whose
resistance, with the current perpendicular to the planes of the layers
(CPP), can change by more than 100\% under the application
of a magnetic field. Until recently N-S and F-S nanostructures
have been studied in isolation, but during recent months, a number of
experiments have demonstrated that N-F-S hybrids exhibit a range of novel
features \cite{sand,petr,upa,upa1}, including a long-range superconducting
proximity effect in the F-material, extending over length scales for an excess of 
the magnetic length 
$\sqrt{\frac{\hbar D}{E_{\mathrm ex}}}$, where $D$ is the diffusion
coefficient and $E_{\mathrm ex}$ the exchange splitting. 

\begin{figure}
\narrowtext
\epsfysize=5cm
\epsfxsize=7cm
\centerline{\epsffile{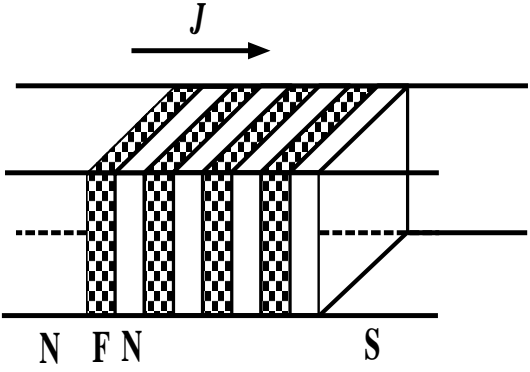}}
\caption{ \label{mul}
Scheme of the system considered: N-F multilayer sandwiched between
a normal (N) contact and a superconducting (S) contact. In the calculation
presented N=Cu, F=Co and S=Pb.}
\end{figure}

The aim of this Letter
is to address a new phenomenon, namely the effect of superconductivity
on CPP-GMR in phase-coherent multilayers of the type (N/F)$_n$S, 
where S is either a superconductor or a normal metal. We shall demonstrate
that as superconductivity is induced in the S-contact (eg by lowering
the temperature) CPP-GMR is almost completely suppressed. This result is
remarkable, since for example CPP-GMR experiments by the
Michigan State University group \cite{mic1,mic2}, already employ 
superconducting contacts. We shall argue that GMR in such experiments is a 
consequence of spin-flip scattering, which if eliminated, would cause a dramatic 
suppression of GMR. To demonstrate the superconductivity-induced suppression
of GMR, we use the method outlined in \cite{noi} to
compute the zero-bias, zero-temperature conductance of the
(Cu/Co)$_n$Pb multilayer sketched in figure \ref{mul}, described
by an spd tight-binding Hamiltonian, with tight-binding parameters
fitted to accurate {\it ab-initio} density functional calculations
\cite{papa}. 
Since we are interested in a phase-coherent nanostructure in which the 
magnetic moments of successive Co layers are either parallel or anti-parallel,
the conductance in the normal state is given by the Landauer formula

\begin{equation}
G_{\mathrm NN}=\frac{e^2}{h}\left(T_\uparrow+T_\downarrow\right)
\;{,}
\label{gnn}
\end{equation}

where $T^\sigma={\mathrm Tr}\;t^\sigma t^{\sigma \dagger}$, with $t$ the 
multi-channel transmission matrix for the structure. In the superconducting
state, equation (\ref{gnn}) is replaced by current-voltage relations derived in 
\cite{colin2}, and re-derived in \cite{colin3,colin4}, 
which in the absence of quasi-particle 
transmission through the superconductor yields

\begin{equation}
G_{\mathrm NS}=\frac{4e^2}{h}R_{\mathrm a}
\;{,}
\label{gns}
\end{equation}

where $R_{\mathrm a}=\;r_{\mathrm a}^\sigma r_{\mathrm a}^{\sigma \dagger}$
is the Andreev reflection coefficient, which for a spin-singlet 
superconductor is independent of the spin $\sigma$ of the incident
quasi-particle. 
In what follows, when the magnetic moments of adjacent Co layers are
aligned (anti-aligned) we denote the conductances by 
$G^{\mathrm F}_{\mathrm NN}$, $G^{\mathrm F}_{\mathrm NS}$
($G^{\mathrm A}_{\mathrm NN}$, $G^{\mathrm A}_{\mathrm NS}$) and therefore
the GMR ratios are given by 
$M_{\mathrm NN}=(G^{\mathrm F}_{\mathrm NN}-G^{\mathrm A}_{\mathrm NN})/
G^{\mathrm A}_{\mathrm NN}$, 
$M_{\mathrm NS}=(G^{\mathrm F}_{\mathrm NS}-G^{\mathrm A}_{\mathrm NS})/
G^{\mathrm A}_{\mathrm NS}$. 

Consider first the case of quasi-ballistic transport in which there is no 
disorder within the layers, nor at the interface, but
the widths of the Co layers are allowed to fluctuate randomly by 1 atomic layer. 
Such a structure is translationaly invariant in the direction 
($\vec{r}_\parallel$)
parallel to the layers, and the Hamiltonian is diagonal in a Bloch basis 
($\vec{k}_\parallel$). Therefore the trace over scattering channels in equation
(\ref{gnn}) and (\ref{gns}) can be evaluated by computing the scattering matrix at 
separate $k_\parallel$ points. Figure 2a shows results
for the GMR ratio in the normal and
superconducting states, obtained by summing over $5\cdot10^3$ $k_\parallel$
points, and clearly demonstrates a dramatic superconductivity-induced suppression
of GMR. Figure 2b and 2c show results for the individual conductances
(note the difference in scales) and demonstrate that the GMR ratio 
$M_{\mathrm NS}$ 
is suppressed because $G_{\mathrm NS}^{F}$ is drastically reduced compared
with $G_{\mathrm NN}^{F}$.

To understand this effect, consider the simplest possible model of
spin-dependent boundary scattering shown in figure \ref{cartoon}, which
in the limit of delta-function F layers, reduces to the model used to
describe the N-F-S experiment of \cite{upa1}. Fig 3a (3b)
shows a cartoon of a majority (minority) spin, scattering from a series of 
potential barriers in successive aligned F layers. Since the minority spins see
the higher barrier, one expects $T^{\mathrm F}_{\downarrow}<T^{\mathrm F}_{\uparrow}$.
Figures 3c and 3d show the scattering potentials for anti-ferromagnetically
aligned layers, for which 
$T^{\mathrm A}_{\uparrow}=T^{\mathrm A}_{\downarrow}<T^{\mathrm F}_{\uparrow}$.
For such an ideal structure, GMR arises from the fact that 
$T^{\mathrm F}_{\uparrow}\gg T^{\mathrm F}_{\downarrow}$ and
$T^{\mathrm A}_{\uparrow}$.
In the presence of a single superconducting contact
this picture is drastically changed. For ferromagnetically aligned layers, 
figure 3e shows an incident majority electron scattering from a series of low barriers, 
which Andreev reflects as a minority hole and then scatters from a series of high 
barriers (figure 3f). 
The reverse process occurs for an incident minority electron, 
illustrating the rigorous result that the Andreev reflection coefficient
is spin-independent. Figures 3g and 3h illustrate
Andreev reflection in the anti-aligned state. The crucial point illustrated by these 
sketches is that in presence of a S contact for both the aligned (figures 3e and 3f)
and anti-aligned (figures 3g and 3h) states the quasi-particle
scatters from N (=4 in the figures) high barriers and N (=4)
low barriers and therefore at the level of a classical resistor model, one expects 
$G_{\mathrm NS}^{\mathrm F}\approx G_{\mathrm NS}^{\mathrm A}$.

\begin{figure}
\narrowtext
\epsfysize=5cm
\epsfxsize=11cm
\centerline{\epsffile{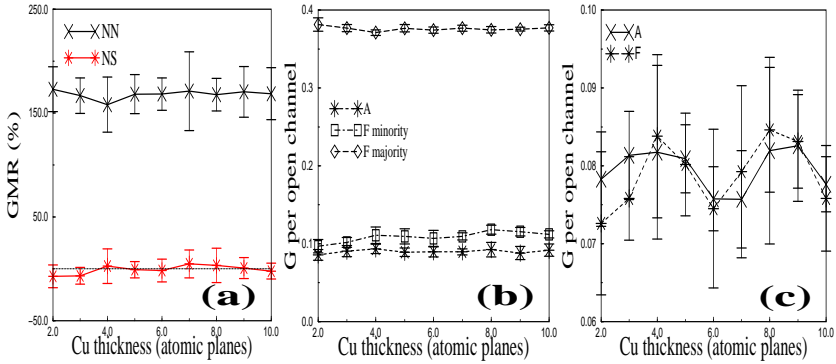}}
\caption{ \label{gmr}
GMR (a) and spin conductances of Cu/Co multilayers as a function of the Cu
thickness, in the case of normal contacts (b)
and with one N contact replaced by a 
superconducting (c) contact.
The system is disorder-free in the layers, but
the Co thickness is allowed to randomly fluctuate by 1 atomic plane.
Every point corresponds to the mean of 10 different
configurations.
The error bars denote the standard deviation of the mean.}
\end{figure}

\begin{figure}
\narrowtext
\epsfysize=8cm
\epsfxsize=7cm
\centerline{\epsffile{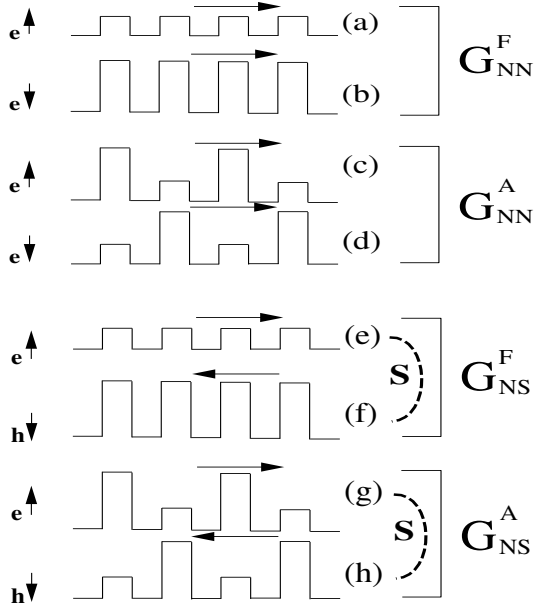}}
\caption{ \label{cartoon}
Cartoon of the different scattering processes. Figures 3a, 3b, 3c and 3d
describes the transmission of spin electrons $e^{\uparrow(\downarrow)}$
in a NN system. Figures 3e, 3f, 3g and 3h describe the NS case. Note that 
in the F case
a majority (minority) spin electron $e^{\uparrow}$ ($e^{\downarrow}$)
is Andreev reflected as a minority 
(majority) hole $h^{\downarrow}$ ($h^{\uparrow}$). In the 
anti-aligned (A) case
the path of the incoming electrons and outcoming holes 
is identical for both spins. The total number of large barriers
is the same in the A and F case, and this produces GMR
suppression.}
\end{figure}

Of course the rigorous results of figure \ref{gmr}, obtained using an
spd Hamiltonian with 36 orbitals per atomic site (spd$\times$2 for
spin $\times$2 for particle-hole degrees of freedom) go far beyond
this heuristic argument. In the case of
aligned or anti-aligned F-layers the problem involves two independent
spin fluids and therefore the Hamiltonian is block-diagonal
with 18 orbitals per site. The Hamiltonian used to obtain these results is of 
the form

\begin{equation}
H_{\mathrm spd}=H_{\mathrm L}+H_{\mathrm LM}+H_{\mathrm M}+H_{\mathrm MR}+
H_{\mathrm R}
\;{,}
\label{ham}
\end{equation}

where $H_{\mathrm L}$ ($H_{\mathrm R}$) describes a semi-infinite left-hand 
(right-hand) crystalline lead, $H_{\mathrm LM}$ ($H_{\mathrm MR}$) is the
coupling matrix between surface orbitals on the left (right) lead and the left (right)
surface of the magnetic multilayer and $H_{\mathrm M}$ is the
tight-binding Hamiltonian describing the multilayer. Consider first the retarded
Green's function $g=(E-H_{\mathrm L}-H_{\mathrm R}+i0^+)^{-1}$
of the two decoupled semi-infinite leads. If the surfaces of the leads each
contain M
atoms, then $H_{\mathrm LM}$ and $H_{\mathrm MR}$ are $36M\times 36M$ matrices
and the portion of $g$ involving only matrix elements between orbitals on the left and right
lead surfaces is a $(2\times 36M)\times(2\times 36M)$ block diagonal
matrix $g^{\mathrm S}$ whose matrix element $g^{\mathrm S}_{ij}$
vanish for $i,j$ belonging to different leads. Using a semi-analytic form for 
$g^{\mathrm S}$ derived in \cite{noi}, the surface Green function $G^{\mathrm S}$
for the leads plus multilayer can be computed by first recursively decimating the
Hamiltonian $H_{\mathrm LM}+H_{\mathrm M}+H_{\mathrm MR}$
to yield a $(72M \times 72M)$ matrix of couplings $\tilde{H}_{\mathrm M}$
between surface orbitals of the leads, and then computing the inverse

\begin{equation}
G^{\mathrm S}=[(g^{\mathrm S})^{-1}-\tilde{H}_{\mathrm M}]^{-1}
\;{.}
\label{GS}
\end{equation}

Once the full surface Green's function $G^{\mathrm S}$ is known, the scattering
matrix elements between open scattering channels are obtained 
using a generalized Fisher-Lee relations \cite{colin4}.
For the calculation of figure \ref{gmr}, involving fcc crystalline leads 
aligned along the (110) direction the spd Hamiltonians for the bulk materials
are known \cite{papa}, but the hopping parameters at the interfaces
between different materials are not currently available. As a simplest
approximation, these surface couplings were chosen to be the geometric
mean of their bulk values.

Despite our use of a highly efficient
recursive Green's function technique
to exactly evaluate the scattering matrix of a multilayer, currently
available computing resources restrict such a calculation to systems with translational
invariance parallel to the planes. To demonstrate that the suppression of CPP-GMR
is a generic feature of N-F-S hybrids and to
study the effect of elastic impurity
scattering, we now examine a reduced two band (s-d) model with a Hamiltonian matrix

\begin{equation}
H=
\left(
\begin{array}[4]{rr}
\underline{\underline{H_o}}-\underline{\underline{h}} & \underline{\underline{\Delta}}\\
\underline{\underline{\Delta}}^* & -\underline{\underline{H_o}}^*+\underline{\underline{h}}\\
\end{array}
\right)
\;{.}
\label{ham1}
\end{equation}

In this model, $h_{ij}^{\alpha\beta}=h_{i}\delta_{ij}\delta_{\alpha\beta}
\delta_{\alpha\mathrm{d}}$ with $h_i$ 
the exchange splitting on site
$i$ for the d orbital, 
which vanishes if $i$ belongs to a N or S layer and
is of magnitude $h$ if $i$ belongs to a F layer; 
$\Delta_{ij}^{\alpha\beta}=\Delta_{i}\delta_{ij}\delta_{\alpha\beta}$ 
where the superconducting order parameter $\Delta_{i}$
vanishes if $i$ belongs to a N or F layer and equals $\Delta$ if $i$ belongs to the 
S region (s-wave superconductivity);
$(H_o)_{ij}^{\alpha\beta}=\epsilon_i^\alpha\delta_{\alpha\beta}$ 
for $i=j$, $-\gamma^{\alpha\beta}$ for $i,j$ nearest neighbors
and $(H_o)_{ij}^{\alpha\beta}=0$ otherwise. 
Note that this model is the minimal model including
the possibility of s-d interband scattering, that has been shown to play a crucial
r\^ole in describing the scattering properties of a transition metal multilayer
\cite{noi}. $\epsilon_i^\alpha$ is chosen to be a random number, uniformly
distributed between $\epsilon^\alpha-w/2$ and $\epsilon^\alpha+w/2$.
We choose the parameters of the model to fit the conductance
and the GMR obtained from the spd model for Cu/Co \cite{noi}, namely (all quantities
are expressed in eV)
$\epsilon^{\mathrm s}_{\mathrm Cu}=-7.8$, $\epsilon^{\mathrm d}_{\mathrm Cu}=-4.0$,
$\gamma^{\mathrm ss}_{\mathrm Cu}=2.7$, $\gamma^{\mathrm dd}_{\mathrm Cu}=0.85$,
$\gamma^{\mathrm sd}_{\mathrm Cu}=1.1$,
$\epsilon^{\mathrm s}_{\mathrm Co}=-4.6$, $\epsilon^{\mathrm d}_{\mathrm Co}=-2.0$,
$\gamma^{\mathrm ss}_{\mathrm Co}=2.7$, $\gamma^{\mathrm dd}_{\mathrm Co}=0.85$,
$\gamma^{\mathrm sd}_{\mathrm Co}=0.9$, $h=1.6$,
$\Delta=10^{-3}$, $w=0.6$. 

\begin{figure}
\narrowtext
\epsfysize=5cm
\epsfxsize=11cm
\centerline{\epsffile{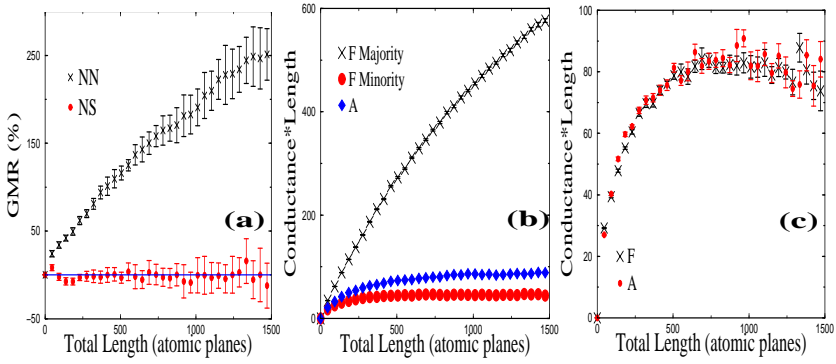}}
\caption{ \label{disor}
GMR (a) and spin conductances of Cu/Co multilayers
calculated with the s-d two bands
model, in the case of normal contacts (b) 
and with one N contact replaced by a 
superconducting (c) contact. The Co and Cu thicknesses are
fixed and are respectively 15 and 8 atomic planes. Every point on the graph
corresponds to an additional double bilayer Co/Cu/Co/Cu.
The on-site energy
fluctuates randomly according with a normal distribution of
width $w=0.6$, and the error bars are the
standard deviation of the mean over 10 random configurations.
The unit cell is a square with $5\times 5$ atomic sites, and
we consider 25 $k-$points in the Brillouin zone.
The horizontal line denotes GMR=0.}
\end{figure}

Figure 4a shows results for the GMR ratios $M_{\mathrm NN}$
and $M_{\mathrm NS}$ and demonstrates that the suppression of CPP-GMR
by superconductivity survives in the presence of disorder.
We have investigated a range of higher disorders and system sizes and
find superconductivity-induced GMR suppression in all cases. The disorder used
in figure \ref{disor} has been chosen to illustrate an additional novel feature,
not so-far discussed in the literature, namely that ballistic majority spins can 
co-exist with diffusive minority spins. For a strictly ballistic structure,
the conductance $G$ is almost independent of length $L$ and the product
$G\cdot L$ varies linearly with $L$. This
behavior occurs for the majority spin in the N-results of figure 4b. In
contrast, for minority spins the product  
$G_{\mathrm NN}^{\mathrm F\downarrow}\cdot L$ exhibits a plateau for 
$L\ge 500$ AP, which is characteristic of diffusive behavior. The same plateaus are also 
present in the product $G_{\mathrm NN}^{\mathrm A}\cdot L$ and in the
curves of $G_{\mathrm NS}^{\mathrm A}\cdot L$ and
$G_{\mathrm NS}^{\mathrm F}\cdot L$ shown in figure 4c.

In summary we predict that the presence of a single superconducting contact
destroys the sub-gap CPP-GMR of phase-coherent magnetic multilayer.
This arises because superconductivity suppresses transport in the majority
sub-band in the ferromagnetic alignment, but causes little effect 
in the antiferromagnetically aligned state. This drastic
reduction in $G^{\mathrm F}_{\mathrm NS}$ compared with
$G^{\mathrm F}_{\mathrm NN}$ is itself a remarkable superconductivity-induced
magneto resistance effect. 
This suppression will be lifted at high biases and finite temperatures, where transport 
occurs via both Andreev reflection and quasi-particle transmission. Furthermore
the presence of spin-flip scattering at the superconducting interface will destroy
this effect, because if the spin of an Andreev reflected hole is flipped by
such process, before it traverses the multilayer, only the contribution
to GMR from layers within a spin-flip scattering length of the N/S interface
is suppressed. 

{\bf Acknowledgments}: This work is supported by the EPSRC, the EU
TMR Programme and the DERA.

\end{multicols}

\end{document}